\documentclass[prd,twocolumn,amsmath,amssymb,nofootinbib,preprintnumbers,balancelastpage]{revtex4}
\usepackage{subfigure}
\usepackage{graphicx}
\usepackage{cancel}
\usepackage{amssymb}
\usepackage{textcomp}
\usepackage{amsmath}
\usepackage{bm}
\usepackage{times}
\usepackage{epsfig}
\usepackage{color}
\def\beq{\begin{equation}}
\def\eeq{\end{equation}}
\def\bea{\begin{eqnarray}}
\def\eea{\end{eqnarray}}
\def\bean{\begin{eqnarray*}}
\def\eean{\end{eqnarray*}}

\usepackage{hyperref}
\hypersetup{
    pdfnewwindow=true,     
    colorlinks=true,       
    linkcolor=black,       
    citecolor=blue,        
    filecolor=blue,        
    urlcolor=blue          
}

\begin{document}
\title{\Large {B and L at the SUSY Scale, Dark Matter and R-parity Violation \vspace{0.5cm}}}
\author{Jonathan M. Arnold}
\affiliation{California Institute of Technology, Pasadena, CA 91125, USA}
\author{Pavel Fileviez P\'erez}
\affiliation{Particle and Astro-Particle Physics Division \\
Max Planck Institute for Nuclear Physics {\rm{(MPIK)}} \\
Saupfercheckweg 1, 69117 Heidelberg, Germany}
\author{Bartosz Fornal}
\affiliation{California Institute of Technology, Pasadena, CA 91125, USA \,}
\author{Sogee Spinner}
\affiliation{Department of Physics, David Rittenhouse Laboratory \\
209 South 33rd Street, University of Pennsylvania, \\
Philadelphia, PA 19104-6395, USA \\}
\begin{abstract}
We present a simple theory where baryon and lepton numbers are spontaneously broken at the supersymmetry scale. In this context R-parity must be spontaneously broken but the theory still contains a stable field which can play the role of the cold dark matter of the Universe. We discuss the spectrum of the theory, the properties of the dark matter candidate and the predictions for direct detection experiments. This theory provides a concrete example of exotic supersymmetric signatures associated with having the simultaneous presence of R-parity violating and missing energy signals at the Large Hadron Collider.
\end{abstract}
\maketitle
\section{Introduction}
The Minimal Supersymmetric Standard Model (MSSM) is considered one of the most appealing candidates for physics beyond the Standard Model. While the recent results from the Large Hadron Collider (LHC)  have set serious constraints on the masses of the supersymmetric particles, if one suspects that new physics exists at an LHC accessible scale, an MSSM-like theory still highly recommends itself as a candidate theory.

Despite its various appealing properties, the MSSM poses a challenge for proton stability. This is because it introduces two separate sets of operators which induce proton decay: tree-level terms, which separately violate baryon and lepton number, and non-renormalizable terms which individually violate both baryon and lepton number. The first of these are
\begin{displaymath}
	\hat{L}\hat{H}_u, \ \hat{L}\hat{L}\hat{e}^c, \ \hat{Q}\hat{L}\hat{d}^c, {\rm{and}}  \  \hat u^c \hat d^c \hat d^c,
\end{displaymath}
where the first three operators violate lepton number and the last baryon number. Any combination of the first three operators with the last one leads to rapid proton decay. Their absence is typically explained by invoking R-parity, an \textit{ad hoc} discrete symmetry defined as $R = (-1)^{3(B-L) + 2S}$, which forbids all of these terms. However, the fate of such operators is most simply divined from models of gauged $B-L$. The most minimal of such models lead to lepton number violating R-parity violation (and therefore no tree-level proton decay)~\cite{Barger:2008wn} but R-parity conserving models are also possible~\cite{Aulakh:1999cd, Aulakh:2000sn, Babu:2008ep}. Regardless of the type of $B-L$ model, the second type of proton decay inducing operators exist. These are non-renormalizable operators which conserve $B-L$ but violate $B$ and $L$ separately, \textit{e.g.}
\begin{displaymath}
\hat{Q}\hat{Q}\hat{Q}\hat{L}/\Lambda, {\rm{and}}  \  \hat{u}^c \hat{u}^c \hat{d}^c \hat{e}^c/\Lambda.
\end{displaymath}
Despite the suppression in these terms due to the scale of new physics, the bounds on proton decay are strong enough to motivate a mechanism for suppressing them. See~\cite{Nath:2006ut} for a review of proton decay.

Recently, a simple theory for the spontaneous breaking of local baryon and lepton numbers has been proposed in Ref.~\cite{Duerr:2013dza}. In this context one can define an anomaly free theory using fermionic leptoquarks which have both baryon and lepton number charges. Furthermore, even after symmetry breaking, the lightest leptoquark is stable due to a remnant $Z_2$ symmetry and can therefore be a dark matter candidate. See also Refs.~\cite{FileviezPerez:2010gw,FileviezPerez:2011pt,Duerr:2013lka} for similar studies. This idea can be applied in the context of supersymmetric theories to establish not only the origin of the R-parity violating terms, as in the $B-L$ models, but also determine the fate of the non-renormalizable terms which violate $B$ and $L$ separately. 

In this paper we investigate an extension of the MSSM where the local baryonic and leptonic symmetries are spontaneously broken at the supersymmetry scale.
We find that the minimal model predicts that R-parity must be spontaneously broken in the MSSM sector (leading only to lepton number violation). Despite the breaking of R-parity, the remnant $Z_2$ symmetry from the breaking of the baryonic and leptonic symmetries ensures that the lightest leptoquark is stable and may be a candidate for the cold dark matter of the universe. We investigate the spectrum of the theory and the predictions for dark matter direct detection. This article is organized as 
follows: In section II we discuss the model with local $B$ and $L$ symmetries, in Section III we discuss the possible dark matter candidates and the predictions for dark matter experiments. Finally, we summarize our results in Section IV.
\section{Spontaneous Breaking of $B$ and $L$}
In order to define a theory for local baryon and lepton numbers we use the gauge group,
\bean
SU(3)_C \otimes SU(2)_L \otimes U(1)_Y \otimes U(1)_B \otimes U(1)_L\ .
\eean
An anomaly free theory can be achieved by adding the following new leptoquark fields with $B$ and $L$ numbers~\cite{Duerr:2013dza}:
\begin{eqnarray}
\label{Psi}
\hat{\Psi} &\sim& (1,2,-1/2,B_1,L_1) \ , \ \
\hat{\Psi}^c \sim (1,2,1/2,B_2,L_2) \ , \nonumber\\
\hat{\eta}^c &\sim & (1,1,1,-B_1,-L_1) \ , \ \
\hat{\eta} \sim (1,1,-1,-B_2,-L_2) \ , \nonumber\\
\hat{X}^c  &\sim&  (1,1,0,-B_1,-L_1)\ ,   \  {\rm{and}} \ \  \hat{X}  \sim  (1,1,0,-B_2,-L_2) \ .\nonumber
\end{eqnarray}
Notice that these fields are vector-like with respect to the SM transformations. The anomalies can be cancelled for any values of $B_i$ and $L_i$ ($i=1,2$) which satisfy the conditions
\bea
B_1+B_2=-3 \ , \  {\rm{and}} \ \ L_1+L_2=-3 \ .
\eea
In order to generate masses for the new fields and for symmetry breaking we need the chiral superfields,
\begin{eqnarray}
	\hat S_1   \sim  (1,1,0,3,3)\ ,
\  {\rm{and}} \ \hat S_2   \sim  (1,1,0,-3,-3)\ .\nonumber
\end{eqnarray}
Therefore, the superpotential of this theory is given by
\begin{eqnarray}
{\cal{W}}_{\rm{BL}}&=&{\cal{W}}_{\rm{RpC}} \ + \ {\cal{W}}_{\rm{LB}},
\end{eqnarray}
where
\begin{eqnarray}
{\cal{W}}_{\rm{RpC}}&=&Y_u \hat{Q} \hat{H}_u \hat{u}^c + Y_d \hat{Q} \hat{H}_d \hat{d}^c + Y_e \hat{L} \hat{H}_d \hat{e}^c \nonumber \\
&+&  Y_\nu \hat{L} \hat{H}_u \hat{\nu}^c + \mu \hat{H}_u \hat{H}_d,
\end{eqnarray}
contains the R-parity conserving terms present in the MSSM (plus a Yukawa coupling for the neutrinos, $Y_\nu$), and
\begin{eqnarray}
{\cal{W}}_{\rm{LB}}&=&
	Y_1 \hat{\Psi} \hat{H}_d \hat{\eta}^c
	+ Y_2 \hat{\Psi} \hat{H}_u \hat{X}^c
	+ Y_3 \hat{\Psi}^c \hat{H}_u \hat{\eta}
	+ Y_4 \hat{\Psi}^c \hat{H}_d \hat{X} \nonumber
	\\
	&&\hspace{-8mm}+ \lambda_1 \hat{\Psi}  \hat{\Psi}^c \hat{S}_{1}
	+ \lambda_2 \hat{\eta}  \hat{\eta}^c \hat{{S}}_{2}
	+ \lambda_3 \hat{X} \hat{X}^c \hat{{S}}_{2}
	+ \mu_{BL} \hat{S}_{1} \hat{{S}}_{2},
\end{eqnarray}
is the superpotential of the leptoquark sector needed for anomaly cancellation. Of course, because of the conservation of $B$ and $L$, both the R-parity violating terms and the non-renormalizable terms leading to proton decay are forbidden. Notice that when $B_1=B_2$ and $L_1=L_2$ we can have Majorana masses for the $\hat{X}$ and $\hat{X}^c$, 
but we stick to the general case where the quantum numbers are different.

An interesting consequence of the leptoquark sector is the presence of a $Z_2$ symmetry once $S_1$ and $S_2$ acquire a VEV. Under this symmetry, all leptoquarks are odd: $\Psi \to - \Psi$, $\Psi^c \to - \Psi^c$, $\eta \to - \eta$, $\eta^c \to - \eta^c$, $X \to - X$ 
and $X^c \to - X^c$. The consequence of this is that the lightest leptoquark is stable (must be neutral) and therefore a  dark matter candidate.
\subsection{Symmetry Breaking and Gauge Boson Masses}
Symmetry breaking in the baryon and lepton number sector proceeds through the following scalar potential:
\begin{eqnarray}
V &=& \left( M_{1}^2 + |\mu_{BL}|^2 \right) |S_{1}|^2 + \left(M_{2}^2 + |\mu_{BL}|^2 \right) |{S}_{2}|^2 \nonumber \\
& + &  M_{\tilde \nu^c}^2 | \tilde \nu^c |^2 + \frac{9}{2} g_B^2 \left(|S_{1}|^2 -  |{S}_{2}|^2 \right)^2\nonumber \\
& + & \frac{1}{2} g_L^2 \left(3 |S_{1}|^2 \!-\!  3 |{S}_{2}|^2 \!-\! | \tilde \nu^c |^2 \right)^2 \nonumber \\
& - & \left( b_{BL} S_{1} {S}_{2} \!+\! {\rm{h.c.}}\right),
\end{eqnarray}
where $M_1$, $M_2$ and $M_{\tilde{\nu}^c}$ are the soft terms for the scalar fields $S_1$, $S_2$ and $\tilde{\nu}^c$, respectively.
Here $b_{BL}$ is the bilinear term between $S_1$ and $S_2$ and we define the vacuum expectation values (VEVs) as
\begin{eqnarray}
\sqrt 2 S_{1}&=& v_{1} + h_{1} + i a_{1}, \\
\sqrt 2 {S}_{2}&=&  {v}_{2} + {h}_{2}+ i {a}_{2}, \\
\sqrt 2 \tilde \nu^c &=& v_R + h_R +  i a_R.
\end{eqnarray}
The squared mass matrix for the new gauge bosons can be written as
\begin{equation}
	\mathcal{M}_{Z'}^2 = 9
	\begin{pmatrix}
		g_B^2 (v_1^2+v_2^2)
		&
		g_B g_L (v_1^2+v_2^2)
		\\
		g_B g_L (v_1^2+v_2^2)
		&
		g_L^2 (v_1^2+v_2^2) + \frac{1}{9} g_L^2 v_R^2
	\end{pmatrix},
\end{equation}
which has a zero determinant if $v_R = 0$; note that this cannot be modified even in the case where $\left< X \right> \neq 0$. This is a consequence of the fact that when  $S_{1}$ and $S_{2}$ acquire VEVs the symmetry group $U(1)_B \otimes U(1)_L$ is broken to $U(1)_{B-L}$. The $B-L$ symmetry can only be broken by the VEV of the right-handed sneutrino as in Ref.~\cite{Barger:2008wn}. Therefore, we conclude that
\begin{center}
{\textit{R-parity must be spontaneously broken  in this theory !}}
\end{center}
However, it is only lepton number violating R-parity violation and therefore the proton remains safe.

The minimization conditions read as
\begin{eqnarray}
0 &=& \left(M_{1}^2 +  |\mu_{BL}|^2 \right) - b_{BL} \frac{v_2}{v_1} + \frac{9}{2} g_B^2 (v_1^2 - v_2^2) \nonumber \\
&+ & \frac{3}{2} g_L^2 (3v_1^2 - 3v_2^2 - v_R^2)
 \ , \\
0 &=& \left(M_{2}^2 +  |\mu_{BL}|^2 \right) - b_{BL} \frac{v_1}{v_2} - \frac{9}{2} g_B^2 (v_1^2 - v_2^2) \nonumber \\
& - & \frac{3}{2} g_L^2 (3v_1^2 -3 v_2^2 - v_R^2)
 \ , \\
 0 &=& M_{\tilde \nu^c}^2 - \frac{1}{2} g_L^2 \left(3 v_1^2 - 3 v_2^2 -v_R^2 \right),
\end{eqnarray}
and can be reformulated as,
\begin{eqnarray}
v_R^2 &=& \frac{2}{g_L^2} \left[-M_{\tilde \nu^c}^2+\frac{3}{2}g_L^2 \left( v_1^2 -  v_2^2\right)\right] , \\
\sin(2\gamma) &=& \frac{2b_{BL}}{M_1^2+M_2^2+2|\mu_{BL}|^2} \ ,
\end{eqnarray}
where we have defined
\begin{align}
	\tan \gamma & = \frac{v_2}{v_1}\ .
\end{align}
One can easily prove that there is no symmetry breaking in the SUSY limit. Therefore, the $B$ and $L$ 
breaking scales are determined by the SUSY scale.
In order to have a potential bounded from below we must satisfy the condition,
\begin{align}
	2 b_{BL} & < M_{1}^2 + M_{2}^2 + 2 |\mu_{BL}|^2,
\end{align}
and in order to break the symmetry we need the condition
\begin{align}	
	b_{BL}^2 & > \left(M_{1}^2 \!+\! |\mu_{BL}|^2 \!-\! \frac{3}{2} g_L^2 v_R^2\right) \left(M_{2}^2 \!+\! |\mu_{BL}|^2 \!+\! \frac{3}{2} g_L^2 v_R^2\right).
\end{align}
The mixing angle between $Z_1$ and $Z_2$ is defined by
\begin{equation}
	\begin{pmatrix}
		Z_B
		\\
		Z_L
	\end{pmatrix}
	=
	\begin{pmatrix}
		\cos \theta_{BL}
		&
		\sin \theta_{BL}
		\\
		-\sin \theta_{BL}
		&
		\cos \theta_{BL}
	\end{pmatrix}
	\begin{pmatrix}
		Z_1
		\\
		Z_2
	\end{pmatrix},
\end{equation}
where $M_{Z_1} < M_{Z_2}$. The eigenvalues for the new neutral gauge bosons are
\begin{equation}
	M_{Z_{1,2}}^2 = \frac
		{1}{2}\left(
			M_{Z_L}^2 + M_{Z_B}^2 \pm
			\sqrt
			{
			\Delta_{BL}^2
			}
		\right),
\end{equation}
where
\bea
\Delta_{BL}^2 &=& \left(M_{Z_L}^2 - M_{Z_B}^2 \right)^2 + 4 g_L^2 M_{Z_B}^4 / g_B^2,
\\
M_{Z_B}^2 &\equiv& 9 g_B^2(v_1^2+v_2^2), 
\\
M_{Z_L}^2 &\equiv& 9 g_L^2 \left(v_1^2+v_2^2+\frac{1}{9}v_R^2\right),
\eea
and the mixing angle is given by
\begin{align}
	\sin (2 \theta_{BL}) = \frac{2 g_B g_L (v_1^2+v_2^2)}
	{
		M_{Z_2}^2 - M_{Z_1}^2
	}.
\end{align}
Note that this produces a $Z_1$ lighter than $Z_2$ only for $M_{Z_B}<M_{Z_L}$. For the opposite case we take 
$\theta_{BL} \to -\theta_{BL}$ and $Z_1 \leftrightarrow Z_2$. 
In the limit $v_R^2 \gg v_1^2+ v_2^2$ the eigenvalues are
\begin{eqnarray}
	M_{Z_1} &\sim& 9 g_B^2 \left(v_1^2+v_2^2\right)\left(1 - 9 \epsilon \right), \\
	M_{Z_2} &\sim &  g_L^2 v_R^2\left(1 + 9  \epsilon \right),
\end{eqnarray}
where $\epsilon \equiv (v_1^2 + v_2^2)/{v_R^2}$ and the mass eigenstates are,
\begin{align}
	Z_1 & = \left(1 - \frac{81}{2} \frac{g_B^2}{g_L^2} \epsilon^2 \right) Z_B - 9 \frac{g_B}{g_L} \, \epsilon \, Z_L \ ,
	\\
	Z_2 & = 9 \frac{g_B}{g_L} \, \epsilon \, Z_B + \left(1 - \frac{81}{2} \frac{g_B^2}{g_L^2} \epsilon^2 \right) Z_L \ .
\end{align}
This is an interesting limit since the lighter $Z_1$ eigenstate is predominately $Z_B$-like and therefore has lower collider bounds~\cite{An:2012ue, Dobrescu:2013cmh} 
compared to a $Z'$ that significantly couples to leptons~\cite{Carena:2004xs}.

Finally, we note that when baryon and lepton numbers are broken at the SUSY scale, one expects operators mediating proton decay. However, in this theory, the proton is stable because baryon number is broken by three units. The least suppressed 
non-renormalizable terms which generate baryon and lepton number violating interactions occur at dimension 14, \textit{e.g.},
\bea
	{\cal{W}}_{14}  &=& \frac{1}{\Lambda^{10}}
		\Big[
			c_1 \hat{S}_1 (\hat{u}^c \hat{u}^c \hat{d}^c \hat{e}^c)^3  + c_2 \hat{S}_1 (\hat{u}^c \hat{d}^c \hat{d}^c \hat{\nu}^c)^3\nonumber\\
		&&\ + \ c_3 \hat{S}_2 (\hat{Q}\hat{Q}\hat{Q}\hat{L})^3
		\Big].
\eea
Due to this large suppression, there is no need to assume a large scale to be in agreement with experiments.
\subsection{Spontaneous R-parity Violation}
As we saw earlier, in order to avoid a long range $B-L$ force, the sneutrino must acquire a VEV. 
The consequences of this are very similar to those in the minimal supersymmetric $B-L$ model~\cite{Barger:2008wn} and we briefly review them here.

The first and most obvious of these consequences is that R-parity is spontaneously broken. This induces a mixing between SUSY and non-SUSY fields with the same quantum numbers: neutralinos with neutrinos, charginos with charged leptons, sneutrinos with neutral Higgs and charged sleptons with charged Higgs. Typically, the most important of these mixings proceeds through the neutrino Yukawa coupling in the superpotential once the right-handed sneutrino acquires a VEV, and one obtains
\begin{equation}
	W \supset \frac{1}{\sqrt 2} Y_\nu v_R  \ell \, \tilde{H}_u,
\end{equation}
which is the so-called bilinear R-parity violating term usually referred to as $\mu'$. This term also induces a VEV for the left-handed sneutrino which leads to various mixing terms of gauge coupling strength such as $\frac{1}{2} g_1 \tilde B \nu v_L$ and $g_L \tilde B_L \nu v_L$, where $\tilde B$ and $\tilde B_L$ are the hypercharge and lepton number gauginos respectively.
The size of R-parity violation is related to the neutrino sector and is therefore small. Phenomenologically, this means that SUSY processes proceed as if R-parity 
is conserved except for the decay of the LSP, which can now decay into SM states. More specifically, SUSY particles are still pair produced. 
For specific decay channels for a given LSP, see for example~\cite{FileviezPerez:2012mj}.

A further interesting consequence is that a sizable VEV can only be realized for one generation of right-handed sneutrinos. This means that lepton number is broken by one unit only in one generation and it is only the corresponding generation of right-handed neutrinos which attains a TeV scale mass; the other two right-handed, or sterile neutrinos, attain sub-eV masses~\cite{Mohapatra:1986aw, Ghosh:2010hy, Barger:2010iv}. This has important consequences for cosmology in the form of dark radiation in the early universe and for neutrino oscillation anomalies.
\section{Dark Matter Candidates}
After symmetry breaking, the lightest leptoquark is stable due to the remnant $Z_2$ symmetry as discussed earlier. This particle must be neutral and could play the role of dark matter. Furthermore, unlike in R-parity conservation, the lightest leptoquark can be either a fermion or a scalar. The best candidates are the $\hat X$ and $\hat X^c$ superfields since they do not couple to the $Z$. In this study we assume the lightest leptoquark is the fermionic component of $\hat X$ and $\hat X^c$, whose Dirac spinor we refer to as $\tilde{X}$, and focus on its properties. It is also interesting to note that because the mass of $\tilde X$ is given by the VEV of $S_2$, it is automatically at the SUSY scale and therefore WIMP-like. This would not be true if its mass was simply a parameter in the superpotential whose magnitude would be arbitrary. Of course, there is a trade off here with the $\mu$-type problem associated with the $\mu_{BL}$ parameter.

The fermionic dark matter candidate can annihilate into two fermions through the neutral gauge bosons present in the theory:
\begin{align}
	\bar{\tilde{X}} \tilde{X} \  \to \  Z_i  \  \to \  \bar{f} f.
\end{align}
The relevant interactions in this case are
\begin{eqnarray}
	- {\cal L} &=& g_B \bar{\tilde X} \gamma^\mu \left( - B_2 P_L + B_1 P_R \right) Z_{B\mu} \tilde{X} \nonumber \\ 
&+ &  g_L \bar{\tilde X} \gamma^\mu \left( - L_2 P_L + L_1 P_R \right) Z_{L\mu} \tilde{X},
\end{eqnarray}
which in the physical basis reads as
\begin{eqnarray}
- {\cal L} &=& g_B\bar{\tilde X} \gamma^\mu \left( C_{11} P_L +  C_{12} P_R \right) Z_{1\mu} \tilde{X} \nonumber \\
 &+ &  g_B\bar{\tilde X} \gamma^\mu \left( C_{21} P_L + C_{22} P_R \right) Z_{2\mu} \tilde{X},
\end{eqnarray}
where
\begin{eqnarray}
 C_{11} &=&  -B_2\cos{\theta_{BL}} + \frac{g_L}{g_B}  L_2\sin{\theta_{BL}}, \\ 
 C_{12} &=&  B_1 \cos{\theta_{BL}}  - \frac{g_L}{g_B} L_1\sin{\theta_{BL}} , \\
 C_{21} &=& - B_2\sin{\theta_{BL}}  - \frac{g_L}{g_B}  L_2\cos{\theta_{BL}}, \\
 C_{22} &=&   B_1\sin{\theta_{BL}} + \frac{g_L}{g_B}  L_1\cos{\theta_{BL}}.
\end{eqnarray}
\\
Assuming the contribution from $Z_1$ dominates, we find an annihilation cross section
\begin{eqnarray}
\sigma v &=& \sum_q \frac{g_B^4 \tilde{C} }{36 \pi s} \frac{\sqrt{1- 4 m_q^2 / s}}{( s-M_{Z_1}^2)^2 + \Gamma_{Z_1}^2 M_{Z_1}^2},
\end{eqnarray}
where
\begin{eqnarray}
\tilde{C}&=& [(C_{11}^2+C_{12}^2) (s+2 m_q^2) (s-M_{\tilde{X}}^2) \nonumber \\
&+ & 6 \, C_{11} C_{12} M_{\tilde{X}}^2 (s+2 m_q^2) ] \cos ^2{\theta_{BL}},
\end{eqnarray}
This cross section is given by
\begin{eqnarray}
(\sigma v)_{NR} &=& \sum_q \frac{g_B^4}{24 \pi} \frac{\sqrt{1- m_q^2/M_{\tilde X}^2}}{(4 M_{\tilde{X}}^2-M_{Z_1}^2 )^2 + \Gamma_{Z_1}^2 M_{Z_1}^2}  \nonumber \\
&& \times C^2 (2 M_{\tilde{X}}^2+m_q^2)\ ,
\end{eqnarray}
in the non-relativistic limit. Here we have defined
\bea
C = (C_{11}+C_{12}) \cos\theta_{BL}\ .
\eea
In the present epoch the energy density of the relic dark matter particles $\tilde{X}$ would be,
\bea
	\Omega_{\tilde{X}} h^2 \simeq \frac{x_f}{\sqrt{g_*}\,\sigma_0 M_P} \, \frac{(1.07\times10^9)}{{\rm GeV}}\ .
\eea
We adopt the value $\Omega_{DM}^{\rm obs} h^2 = 0.1199 \pm 0.0027$~\cite{Ade:2013zuv}. 
\\
The freeze-out temperature $x_f = M_{\tilde X}/T_f$ is then given by,
\begin{eqnarray}
	x_f &=&
		\ln\!\left(\frac{0.038\,g\, M_P M_{\tilde X} \sigma_0}{\sqrt{g_*}}\right) \nonumber \\
		 &-& \frac{1}{2}
		 \ln\! \left[\ln	\!\left(\frac{0.038\,g\, M_P M_{\tilde X} \sigma_0}{\sqrt{g_*}}\right)\right],
\end{eqnarray}
where $g$ is the number of internal degrees of freedom (in our case $g=4$), $g_*$ is the effective number 
of relativistic degrees of freedom evaluated around the freeze-out temperature, $M_P$ is the Planck 
mass, and we use the expansion $\sigma v = \sigma_0 + \sigma_1 v^2$.

The direct detection also proceeds through the $Z_i$:
\begin{align}
\tilde{X} N \  \to \  Z_i \ \to \ \ \tilde{X} N,
\end{align}
and the spin-independent nucleon-dark matter cross section is then given by
\begin{align}
	\sigma_{\rm SI} = \frac{1}{4 \pi} \frac{M_{\tilde{X}}^2 M_N^2}{(M_{\tilde{X}}+M_N)^2}\frac{g_B^4}{M_{Z_1}^4} C^2,
\end{align}
assuming that the dominant contribution comes from the $Z_1$ gauge boson. 
Because both the dark matter annihilation and direct detection proceed through $Z_1$, they are intimately related to each other. Specifically, once one determines the parameters that yield the correct relic density for a given dark matter mass, there are no free parameters left to hide it from direct detection. Keeping this in mind we present the predictions for the direct detection experiments. 

\begin{figure}
 \centering
 \includegraphics[width=1.0\linewidth]{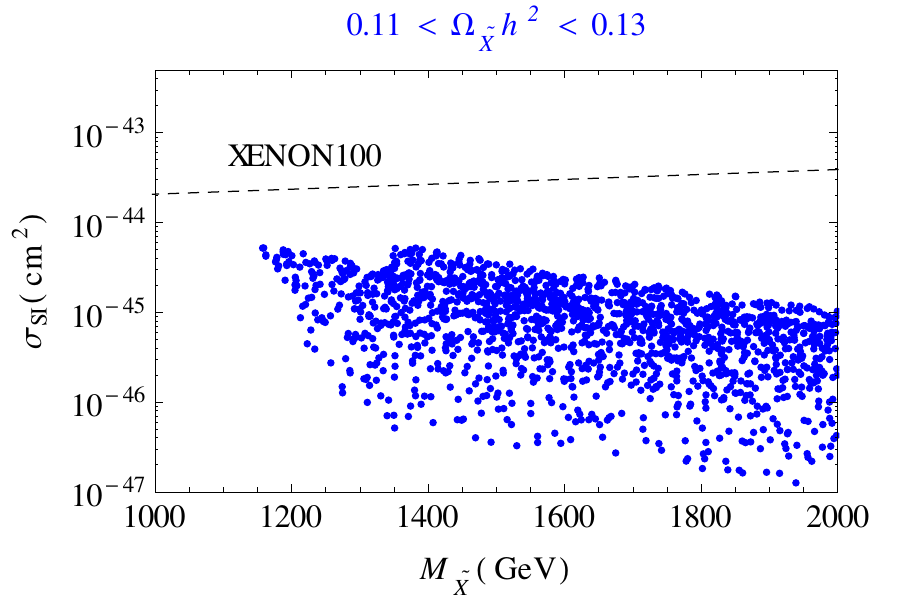}
 \caption{ Predictions for the elastic nucleon-dark matter cross section for different values of the dark matter mass when $0.11 < \Omega_{\tilde{X}} h^2 < 0.13$.}
 \label{DM}
\end{figure}

In Fig.~\ref{DM} we show the values for the spin independent cross section versus the dark matter mass when $C=1$, $0.1 \leq g_B \leq 0.3$, $2.5 \  {\rm{TeV}} \leq M_{Z_1} \leq 5 \ {\rm{TeV}}$, 
and assuming that the relic density is in the range $0.11 < \Omega_{\tilde{X}} h^2 < 0.13$.
\begin{figure}
 \centering
 \includegraphics[width=0.9\linewidth]{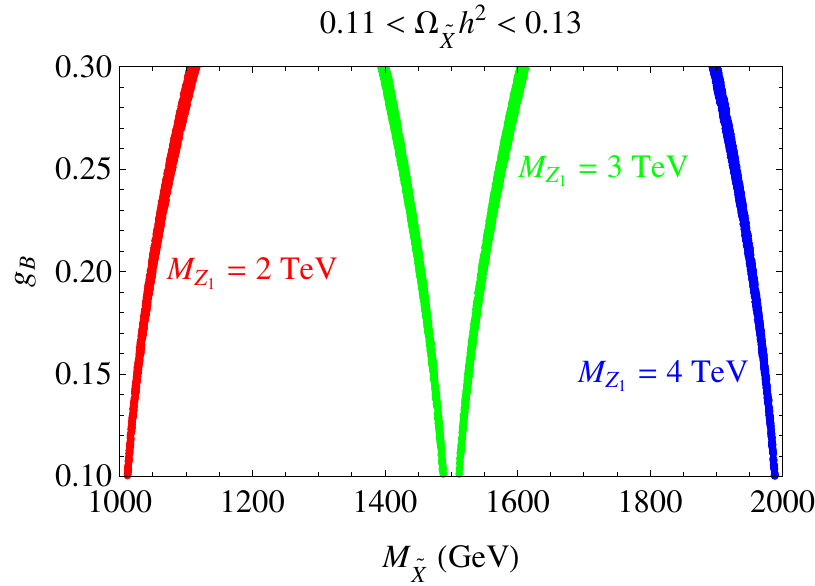}
 \caption{ Allowed values for the gauge coupling and the dark matter mass when $0.11 < \Omega_{\tilde{X}} h^2 < 0.13$ and $M_{Z_1}=2,3,4$ TeV.}
 \label{fig2}
\end{figure}
One can appreciate in Fig.~\ref{DM} that the allowed solutions are below the XENON100 bounds~\cite{Aprile:2012nq}, but could be tested in future dark matter experiments such as XENON1T or LUX.

In Fig. 2 we show some solutions when the mass of new lightest neutral gauge boson is 2, 3, or 4 TeV. 
One can see that there is no need to be very close to the resonance to achieve the required cross section for the relic density. 
\section{Summary and Discussions}
In this article we have presented the simplest supersymmetric extension of the model proposed 
in Ref.~\cite{Duerr:2013dza} where baryon and lepton number are local symmetries.  In this context 
the baryonic and leptonic gauge symmetries are broken at the SUSY scale and the proton is stable.
\\
\\
One of the main predictions of this theory is that R-parity must be spontaneously broken in the MSSM 
sector because the right-handed sneutrino VEVs are needed to break the remnant local $U(1)_{B-L}$ that results from the VEVs of $S_1$ and $S_2$. Even though R-parity is broken, the lightest leptoquark is stable and can be a cold dark matter candidate. The dark matter candidate can be either the spin one-half or spin zero SM singlet leptoquark; we have focused on the former in this paper. It furthermore has baryon and lepton number and therefore couples to the two $Z'$s in the model. 
\\
\\
There are many interesting predictions for the Large Hadron Collider searches in this theory. Since R-parity is broken in the MSSM sector 
we will have lepton number violating signatures at the LHC. For example, one can have exotic channels with four leptons 
and four jets where three of the leptons have the same electric charge~\cite{FileviezPerez:2012mj, Perez:2013kla}. On the other hand there is a stable dark matter candidate in the 
theory which can be produced through the new neutral gauge bosons. Therefore, one can also expect signatures with missing energy 
at the LHC. This theory provides a simple example of very exotic supersymmetric signatures at colliders since one can have the 
simultaneous presence of R-parity violating and missing energy signatures at the LHC.
\\
\\
{\textit{Acknowledgment}}: 
We would like to thank Mark B. Wise for discussions. P. F. P. thanks the theory group at Caltech for support 
and warm hospitality. This paper is funded by the Gordon and Betty Moore Foundation through Grant $\# 776$ to the
Caltech Moore Center for Theoretical Cosmology and Physics. The work of J.A. and B.F. was supported also
in part by the U.S. Department of Energy under contract No. DE-FG02-92ER40701. The work of S. S. is supported 
by the DOE contract DE-AC02-76-ER-03071. 

\end{document}